\documentclass[doublecol]{epl2}
\usepackage{graphicx,amsmath,amssymb,bm,epsf,psfrag,verbatim,subfigure}
\def\prob{p}
\def\pR{p_\mathrm{R}}
\def\bigprob{\mathcal{F}}
\def\order{\mathcal{O}}
\title{Rigidity percolation on the square lattice}
\author{Wouter G. Ellenbroek\inst{1,2} and Xiaoming Mao\inst{1}}
\institute{
\inst{1} Department of Physics and Astronomy, University of Pennsylvania, Philadelphia, PA 19104, USA\\
\inst{2} Department of Applied Physics and Institute for Complex Molecular Systems, Eindhoven University of Technology, P.O. Box 513, NL-5600 MB Eindhoven, The Netherlands
}

\pacs{46.65.+g}{Random media (continuum mechanics)}
\pacs{02.10.Ox}{Graph theory}
\pacs{64.60.ah}{Percolation in phase transitions}

\date{\today}
\abstract{The square lattice with central forces between nearest neighbors
is isostatic with a subextensive number of floppy modes. It can be made
rigid by the random addition of next-nearest neighbor bonds.  This constitutes a
rigidity percolation transition which we study analytically by mapping it to a
connectivity problem of two-colored random graphs. We derive an exact
recurrence equation for the probability of having a rigid percolating cluster
and solve it in the infinite volume limit.  From this solution we obtain the
rigidity threshold as a function of system size, and find that, in the
thermodynamic limit, there is a mixed first-order-second-order rigidity percolation transition at
the isostatic point.}

\begin{document}

\maketitle
\section{Introduction}

Central force rigidity percolation describes how a system of lattice sites
can become rigid by randomly populating the bonds with springs or struts that
can transmit central forces between sites. It features a mechanical critical
point at which an infinite rigid cluster emerges and the system gains
rigidity~\cite{Feng1984,Jacobs1995,Jacobs1996}. Comparing to the analog scalar
problem of connectivity percolation, in which an infinite connected cluster
emerges and the system becomes conducting by populating bonds, rigidity
percolation is a \emph{vector} problem whose basic degrees of freedom are the
$d$-dimensional position vectors of the lattice sites. Another important
character of rigidity percolation is its \emph{long-range} nature, that adding
a bond at one place in the network could affect the rigidity of regions
arbitrarily far away from that bond~\cite{Jacobs1995,Jacobs1996}, which makes
rigidity percolation a challenging problem for theoretical study.  Numerical
simulations on rigidity percolation revealed highly nontrivial physics.  In two
dimensions, numerical simulations using the ``pebble game'' algorithm on
generic networks strongly suggest that the rigidity percolation is second
order~\cite{Jacobs1995,Jacobs1996,Moukarzel1999}.  However the
possibility of a weakly first order transition~\cite{Obukhov1995,Duxbury1999}
is not completely ruled out due to finite size effects in the simulations.  
In three dimensions, numerical
simulations using the ``pebble game'' algorithm found rigidity percolation to
be first order~\cite{Chubynsky2007}.

One special class of rigidity percolation occurs on periodic isostatic
lattices with the nearest-neighbor (NN) bonds already present from the
start~\cite{Souslov2009}. These lattices are at the onset of mechanical rigidity
(isostaticity) because they have equal numbers of degrees of freedom and constraints
in the bulk, which for central forces corresponds to a coordination number $z=2d$,
where $d$ is the dimensionality of the system~\cite{Maxwell1864}. Examples are the
two-dimensional square and kagome lattices and the three-dimensional cubic lattice.
In a finite isostatic lattice, sites on the boundary have coordination number
lower than $2d$, which gives rise to a subextensive number of deformation modes
in which none of the bonds change length. These so-called floppy modes do not
cost any elastic energy, and are extended across the lattice: the system cannot be
macroscopically rigid unless all floppy modes are somehow removed.
To restore rigidity, one can
randomly add some of the next-nearest-neighbor (NNN)
bonds~\cite{Garboczi1985,Obukhov1995,Moukarzel1997,Moukarzel1999,Mao2010}.
This constitutes a special kind of rigidity percolation problem because all
lattice sites are already connected and one only needs to remove a subextensive
number of floppy modes. One therefore expects that a subextensive number of NNN
bonds is enough to provide rigidity.

\begin{figure*}[!t]
\centering
\includegraphics[width=12cm]{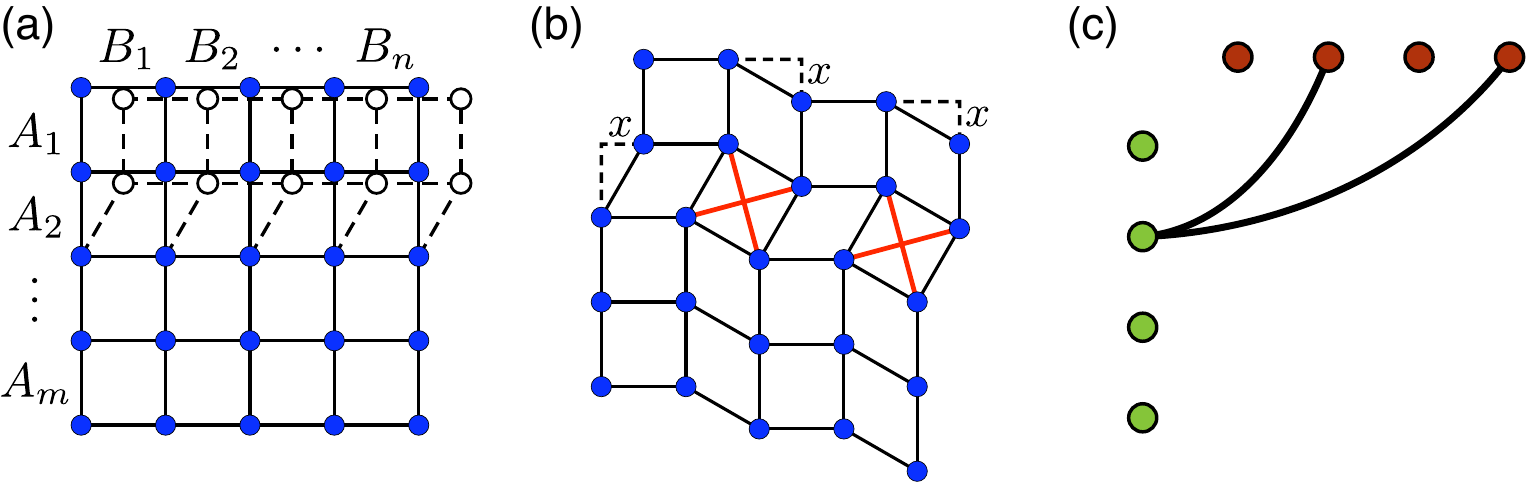}
\caption{(a) The variables $A_i$ and $B_i$ in which we express the zero energy
deformations of the square lattice (before adding NNN bonds) are the relative
displacements of entire neighboring rows and columns, respectively. The effect of
a nonzero $A_2$ is illustrated. (b) Adding a NNN bond at location $(i,j)$ corresponds to constraining
the variables $A_i$ and $B_j$ to be equal, illustrated using $B_2=A_2=B_4=x$. (c) Graph
representation of the network shown in (b), where each variable is represented by a node (light green and dark red for the $A_i$ and $B_i$, respectively)
and each NNN bond by a link between the corresponding nodes.}
\label{fig:vardef}
\end{figure*}

In this Letter, we investigate rigidity percolation on the square lattice,
which has coordination number $z=4=2d$, and is
isostatic~\cite{Souslov2009,Mao2010}. We develop an analytical method to calculate the probability
$\bigprob(L,\prob)$ that a size $L\times L$ square lattice is made rigid by
populating the NNN bonds with probability $\prob$, and thus we arrive at the
rigidity threshold $\pR(L)$ that half of the realizations are rigid, i.e.,
$\bigprob(L,\pR(L))=1/2$.  
Surprisingly, we find that the intuitive argument used in earlier
works~\cite{Obukhov1995,Moukarzel1997,Moukarzel1999}, that having at least one
NNN bond in each row and column should provide rigidity, presents a necessary
but insufficient condition for rigidity of square lattices.
In fact, one can add as few as $L-1$ NNN bonds to an
$L\times L$ square lattice to make the ``one-per-row'' condition satisfied, but
according to the Maxwell counting~\cite{Maxwell1864} there are still $L-2$
floppy modes.

To obtain the rigidity threshold, we map the problem into a
connectivity problem of two-colored (also known as \emph{bipartite}) random graphs with $L-1$ nodes of each
color, for which the number of connected clusters corresponds to the number of
floppy modes in the square lattice. We derive a recurrence relation for the
probability that all nodes of the graph are connected into one cluster,
which corresponds to the probability that the entire square lattice is rigid.
We solve this equation asymptotically in the large $L$ limit, and show that, (i)
for nonzero $\prob$, $\bigprob(L,\prob)\sim 1-2L (1-\prob)^{L}$, (ii) the
rigidity threshold $\pR$ approaches zero when $L\to\infty$ as $\pR(L)\sim \ln
L/L +\order(1/L)$. We confirm these results using numerical simulations.

Consider a regular square lattice with $L$ sites per row or column.  According to
Maxwell~\cite{Maxwell1864}, the number of floppy modes can be found by
subtracting the number of constraints from the number of degrees of freedom
(usually, the global translations and rotations are subtracted as well, leaving
only the nontrivial floppy modes).  Thus the number of floppy modes in the
$L\times L$ square lattice is $2L^2-2L(L-1)-3=2L-3$ which can be understood as follows.  
Each of the rows or columns of square plaquettes can be deformed into
rhombi without changing the shape of the plaquettes in other rows or columns,
as shown in fig.~\ref{fig:vardef}a.  
These deformations correspond to linearly
independent floppy modes of the lattice.  The total number of such rows and
columns of plaquettes is $2L-2$. 
Note that any pure shear deformation can be obtained using linear combinations of these modes,
which is why the shear modulus is zero
~\footnote{Due to anisotropy, strictly speaking, the
square lattice has three elastic moduli, $C_{11}$, $C_{12}$ and $C_{44}$.  With
NN bonds only we have $C_{12}=C_{44}=0$, and in this paper we refer to $C_{44}$ as
shear modulus for convenience~\cite{Ashcroft}.}.
Because the plaquette deformations concern \emph{relative} displacements of
neighboring rows or columns, the space they span does not contain the global translations (but global rotation is included).  

To remove these floppy modes and restore shear rigidity, we can add $2L-3$ NNN
bonds to, for example, the left and the lower boundary plaquettes of the
lattice.  However, for the case of randomly populating each NNN bond with
probability $\prob$, some of the NNN bonds could be redundant, and thus one may
need more than $2L-3$ NNN bonds to restore the shear rigidity. Determining the
rigidity of a disordered network of springs always involves such counting
arguments comparing the number of degrees of freedom to the number of
\emph{independent} constraints, and because of the disorder it is generally
nontrivial to properly keep track of which constraints are truly independent,
which makes rigidity percolation difficult theoretically~\cite{Jacobs1995,Jacobs1996}. 
Below, we will 
derive simple rules for the removal of these floppy modes that make rigidity
percolation on the square lattice tractable.

The spring networks obtained through the addition of
NNN bonds,
while the NN bonds of the square lattice are already present from
the start, are fundamentally different from the classical case of rigidity
percolation on the triangular lattice, where no bonds are present when $p=0$.
Most strikingly, because all sites are already connected, the probability
that a site belongs to the rigid cluster equals one as soon as $\prob$ is large
enough to have a spanning rigid cluster. Thus, in the thermodynamic limit, the
order parameter, defined as the probability that an arbitrarily chosen site belongs to 
the percolating rigid
cluster, jumps from 0 to 1 at the transition, characterizing a first order
transition. On the other hand, the system displays a diverging isostatic length
scale $l^*\sim1/\prob$ and a vanishing shear modulus
$\mu\sim\prob^2$~\cite{Mao2010}, which are characters of a second order
transition. The discontinuous change in the order parameter and
the continuous change in the shear modulus can be heuristically understood by realizing that
the addition of a small number of bonds restores the rigidity of the whole
system, thus these bonds carry all stress applied to the system, resulting in a
vanishingly small shear modulus~\cite{Obukhov1995}.
This feature of a mixed first-order-second-order 
phase transition resembles the jamming transition in packings of
frictionless spheres, which features a jump in the average coordination number
from zero to the isostatic value, as well as power law scalings of the elastic moduli and the isostatic length scale~\cite{Liu1998,O'Hern2003}. The connection
between the square lattice and the jamming transition has been discussed in
Refs.~\cite{Souslov2009,Mao2010}. 

\section{The probability of rigid configurations}
A square lattice of $(m+1)\times (n+1)$ sites with randomly added next
nearest neighbor bonds is rigid if the only zero-energy modes are the global
translations and rotations. To calculate the probability that this is the case,
we consider the space of floppy modes of the original square lattice, which is a
subspace of the space of all modes of the lattice.  In particular, in linear
elasticity the space of the floppy modes is the null space of the dynamical
matrix of the lattice. To exclude the trivial global translational degrees of
freedom, we choose to project this space of floppy modes into the basis of 
deformations of the square plaquettes in each row or column into rhombi,
as discussed earlier and shown in fig.~\ref{fig:vardef}a, or more
precisely, the relative horizontal (vertical) displacements between neighboring
rows (columns).  The dimension of this space is $m+n$ with the global rotation
included.  This leads to a description in terms of relative row-displacements
$A_1,A_2,\ldots,A_m$ and relative column-displacements $B_1,B_2,\ldots,B_n$.

As shown in fig.~\ref{fig:vardef}a, exciting the floppy mode labelled by $A_2$
turns all squares on the corresponding lattice row into rhombi. To make the
system rigid, all floppy modes have to be constrained, so that one needs to
have at least one NNN bond on each row and on each column. However, as will
soon become clear, this necessary condition for rigidity is \emph{not}
sufficient.

Adding a NNN bond to the square at $(i,j)$ constrains it to remain square in
any floppy deformation: it is only allowed to rotate. This amounts to setting
$A_i=B_j$ (with the proper choice of signs). Each NNN bond provides such an
equality, as illustrated in fig.~\ref{fig:vardef}b, in which two NNN bonds
set $B_2=A_2=B_4=x$. The rigidity of the whole lattice would correspond to having
$A_i=B_j=x$ for all $i=1,\ldots,m$ and $j=1,\ldots,n$, with $x$ corresponding to
the amount of global rotation in this case. This suggests a graph representation of realizations of
the rigidity percolation procedure. The nodes represent the variables $A_i$ and
$B_j$, and edges between two nodes represent equalities of the corresponding variables
that arise from the NNN bonds (fig.~\ref{fig:vardef}c). Hence, we obtain a
description of rigidity percolation in terms of a random graph with two types
of nodes (green $A$ nodes and red $B$ nodes), with edges between an $A$-node and a
$B$-node present with probability $\prob$.  
This mapping has been employed before in the context of studying rigidity
in square and cubic structures with added NNN bonds, albeit without the
probabilistic aspect of rigidity percolation~\cite{bolker79}.

Within this description, a rigid lattice is represented by a connected graph,
i.e., a graph in which there is a path along edges between any given pair of
nodes, corresponding to the case of all $A_i=B_j$ with global rotation left as
the only degree of freedom. Note that we use crossbars to represent the effect
of the NNN bonds, because the second NNN bond on a plaquette is redundant, when
only the question ``rigid or not'' is considered. Near the percolation
transition in large systems where $\pR\to0$, it is equivalent to use crossbars with probability
$p$ or to use separate NNN bonds with probability $p/2$, so the resulting scalings are the same.

We now derive the scaling of the probability $\bigprob(m,n,\prob)$ that the
system is rigid, as $m,n\to\infty$. This can be obtained as a corollary from
the results of Pal\'asti~\cite{palasti63}, but we choose to present our own,
more accessible derivation.

For the usual Erd\H os-R\'enyi model~\cite{gilbert59,erdos59,erdos60} for generating
random graphs in which there are $n$ equivalent nodes and each of the
$n(n-1)/2$ edges connecting any two nodes is present with probability $\prob$,
the probability $\bigprob_1(n,\prob)$ that the graph is connected can be
calculated recursively from an expression due to Gilbert~\cite{gilbert59}
\begin{equation}
1-\bigprob_1(n,\prob)=\sum_{k=1}^{n-1} \binom{n-1}{k-1} \bigprob_1(k,\prob) q^{k(n-k)},
\label{onecolor}
\end{equation}
where $q=1-\prob$. Here, the probability that the graph is not connected is
written as the sum over all possible sizes $k$ of the cluster that an
arbitrarily chosen node (labelled as node 1) is part of.  The binomial
coefficient represents the number of ways to choose the other nodes in the
cluster, and the power of $q$ denotes the probability that none of those $k$
nodes are connected to any of the other $n-k$ nodes. Gilbert showed that for
large $n$, $\bigprob_1(n,\prob)\to 1-nq^{n-1}$~\cite{gilbert59}, so the
connectivity threshold for the usual Erd\H os-R\'enyi model is $p_C(L)\sim \ln
n/n$.

We adapt eq.~(\ref{onecolor}) for our two-colored graph in which edges between
differently colored nodes are present with probability $\prob$. Again, we
consider the sum over all the possibilities for the size of the cluster having
$k$ green nodes and $l$ red nodes that an arbitrary green node (to which we
assign the variable $A_1$) is part of:
\begin{equation}
1=\sum_{k=1}^{m}\sum_{l=0}^{n}\binom{m-1}{k-1}\binom{n}{l}\bigprob(k,l,\prob)q^{k(n-l)}q^{l(m-k)}~.
\label{recur2c}
\end{equation}
A recurrence relation for $\bigprob(m,n,\prob)$ can be obtained by moving the term for
$k=m, l=n$ from the sum to the left hand side of the equation.
In the Appendix we show that as $m=n\to\infty$ the
probability approaches
\begin{equation}
\label{limitresult}
\bigprob(n,n,\prob)\approx 1-2nq^n\qquad\mathrm{as~}n\to\infty~.
\end{equation}
For any finite $\prob$ the probability approaches unity, and the value of $\prob$ needed to make half
the realizations rigid approaches zero as (see Appendix)
\begin{equation}
\label{pcritscaling}
\pR(n)\approx\frac{\ln n}{n}+\order(1/n)\qquad\mathrm{as~}n\to\infty~.
\end{equation}
\section{Numerics}
\begin{figure}[t]
\includegraphics[width=8.8cm]{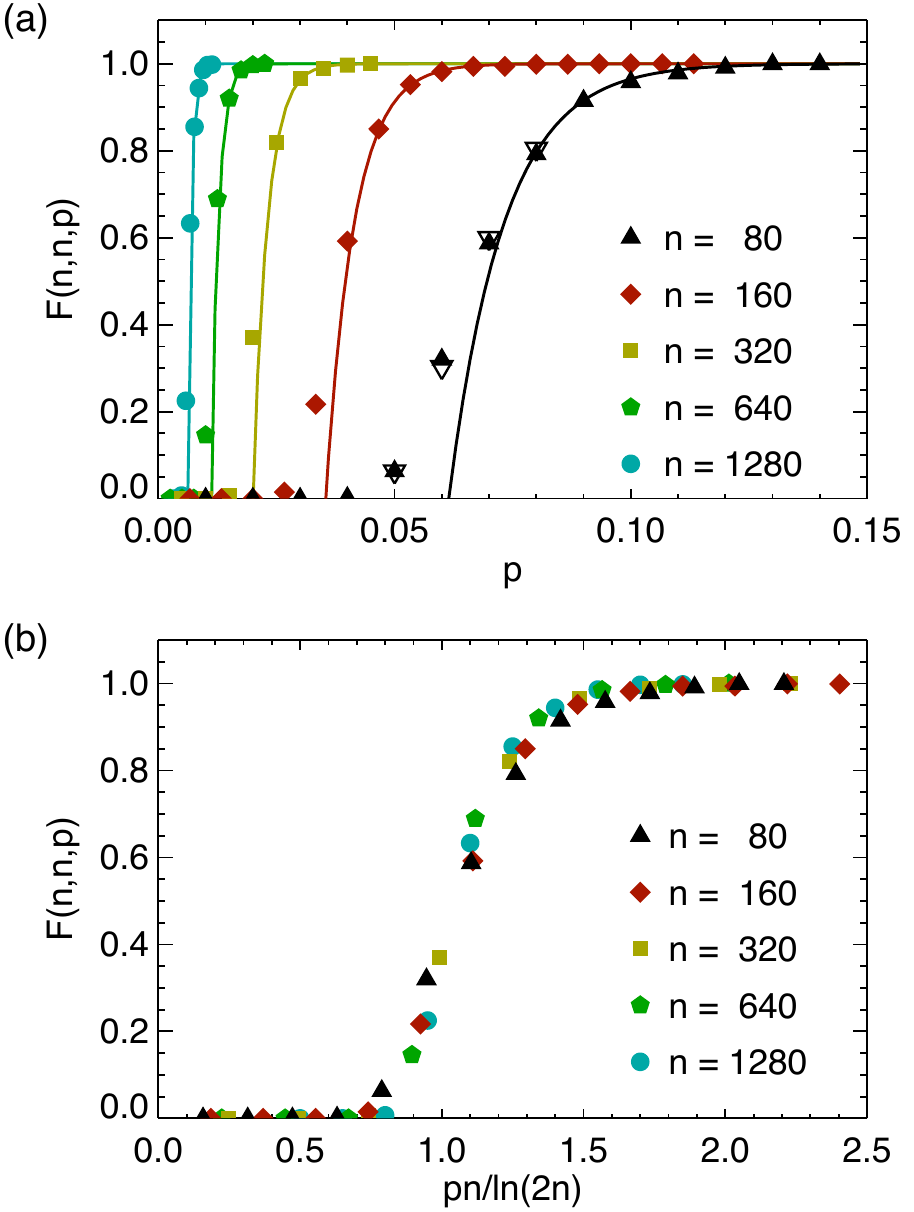}
\caption{(a) Numerically obtained fraction of rigid configurations $\bigprob(n,n,p)$ as a function of
bond-presence probability $\prob$ for various system sizes. The solid lines denote the asymptotic form
$\bigprob=1-2n(1-p)^n$ [eq.~(\ref{limitresult})]. For completeness we included a few
points (open triangles) obtained by numerically evaluating the recurrence relation~(\ref{recur2c}) for $n=80$.
(b) The same data with $p$ rescaled by $\ln(2n)/n$,
according to eq.~(\ref{pcritscaling}).}
\label{fig:numerics}
\end{figure}
For various values of $n$ and $\prob$, we generated 1000 trials of the rigidity percolation procedure and
analyzed the corresponding graphs using an adapted version of the Hoshen-Kopelman cluster labelling
algorithm~\cite{hoshen76}.  We plot the fraction $\bigprob$ of the graphs that are connected, corresponding
to the fraction of rigid networks, in
fig.~\ref{fig:numerics}a. The solid lines in this figure indicate the asymptotic form derived in
eq.~(\ref{limitresult}). Figure~\ref{fig:numerics}b shows the same data, with the $\prob$-axis rescaled by
$\ln(2n)/n$ [the 2 comes from the minimum prefactor of the $1/n$ term in eq.~(\ref{pcritscaling}) --- see Appendix],
clearly showing the scaling of $\pR(n)$.

\section{Discussion}
\begin{figure}[t]
\includegraphics[width=8.6cm]{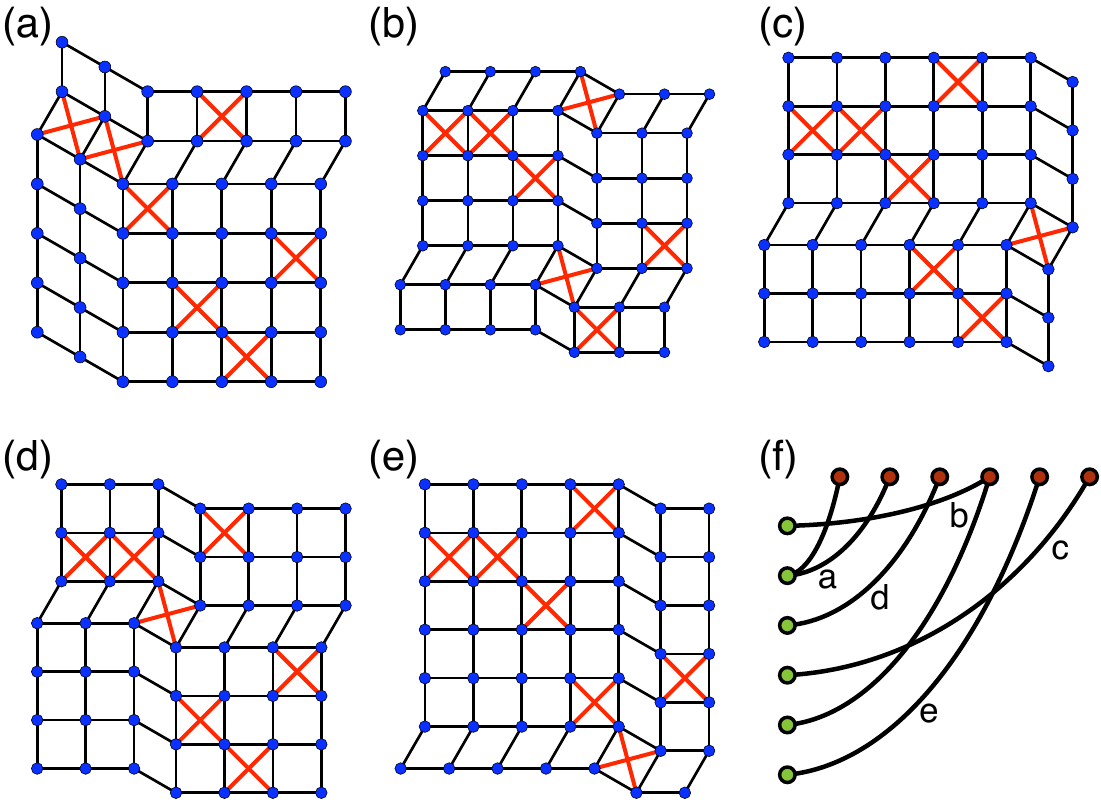}
\caption{Having a crossbar on each row and each column does not guarantee rigidity. (a-e) The five 
floppy modes of an example system. (f) The graph representation of this example system. The five connected clusters
correspond to the five floppy modes in panels a-e of this figure.}
\label{fig:floppyRC}
\end{figure}

The graph picture for rigidity percolation on the square lattice provides a
transparent way to keep track of the effect of added constraints. The intuitive
argument used in earlier works~\cite{Obukhov1995,Moukarzel1997,Moukarzel1999}, that having at
least one NNN bond in each row and column should provide rigidity, corresponds
to not having any isolated nodes with no edges. Clearly this is a weaker
condition than demanding the graph to be connected. This is illustrated
in fig.~\ref{fig:floppyRC}, in which panels (a-e) show the five floppy modes of
a lattice in which the condition of having at least one NNN bond in each row
and column is satisfied. The corresponding graph is shown in
fig.~\ref{fig:floppyRC}f. While counting the number of floppy modes by just looking
at a picture of a lattice is far from trivial, the graph representation gives a
much simpler view in which it is easy to see that there are 5 distinct
connected components, so that the lattice has 5 floppy modes.

Thus, using the weaker ``one-per-row'' condition leads to a structural underestimation of the
rigidity threshold $\pR$. However, this condition does provide the correct
scaling with system size, $\pR\sim\ln L/L$~\cite{Moukarzel1997}. This implies
that the fraction of realizations that satisfy the ``one-per-row''
condition but are not fully connected vanishes as $L\to\infty$, which is consistent
with the known result that just below the transition, the graph contains just one
giant component and some isolated nodes~\cite{saltykov95}.

As an aside, we note that the appearance of this giant component is in itself an
important transition in the theory of random graphs~\cite{erdos60}.  However,
for the system studied in this paper its relevance is limited. Having a giant
component is not enough for rigidity to percolate, because any isolated node
that is not connected to the giant component corresponds to a row or column in
the lattice in which there are no crossbars, so that the two parts of the
system on either side of that row or column can freely shear past each other. Thus the
relevant transition is indeed that to connected graphs, as described in this letter.

The formulation of our approach in terms of the null space of the dynamical matrix
implies that the floppy modes we talk about in principle only need to be floppy up to
linear order, i.e.\ for infinitesimal deformations.  However, 
the row and column displacements in our discussion can be straightforwardly
extended to finite displacements, and thus the mapping to
the graph representation is not limited to linear elasticity. This can be easily seen in
fig.~\ref{fig:floppyRC}, in which the modes have a finite
amplitude but are clearly zero energy modes.

In our discussion the square lattice has been assumed to be a perfect periodic
lattice.  However, ``generic lattices'', in which only the connectivity
(topology) is prescribed but bond lengths are not all equal, are often used in
the discussion of rigidity theory~\cite{Jacobs1996,Connelly2005}. The percolation
problem is still well-defined: whether or not a given structure (a set of
points and rigid bars connecting them) is rigid does not depend on the precise
positions of the points, but only on the connectivity and on the fact that the
points do not have any positional order~\footnote{Genericness of a set of site
positions is usually defined as there not being any nontrivial algebraic
equation (with rational coefficients) relating the coordinates of the
sites.}~\cite{Connelly2005}. This, again, seems to suggest some kind of graph
theoretical approach could be helpful. However, the mapping to the graph
connectivity problem that we used in this paper relies specifically on the
positional order of the sites and hence does not work for the case of generic lattices. The
reason is that some of the clusters of the resulting graph represent internally
inconsistent constraints in this language. These clusters do not correspond to
floppy modes, so that the lattice could become rigid before the graph is
fully connected. The threshold probability $\pR(n)$ for periodic lattices
therefore serves as an upper bound for the $\pR(n)$ for generic lattices, but there is
no obvious lower bound, because the ``one-per-row'' condition that provided the
lower bound on $\pR$ in the perfect square lattice (see appendix) is no longer
a necessary condition for rigidity. Clearly, a more sophisticated approach is
needed to describe generic rigidity percolation. The opportunities that arise
from mapping this problem to so-called \emph{hypergraphs} are currently being
explored~\cite{chentbp}.

In conclusion, we have derived an exact recurrence equation for the
probability $\bigprob(L,L,\prob)$ that a square lattice with given size $L$ and
NNN bonds occupation probability $\prob$ is rigid, and obtained the asymptotic
solution to this equation in the limit of $L\to\infty$.  Our results
unambiguously prove that the rigidity percolation in the square lattice occurs
at $\prob=0$ and is 
\emph{a mixed first-order-second-order transition}: In the thermodynamic limit, on the one hand all lattice sites immediately become part of a percolating rigid cluster as soon as $\prob$ is nonzero; on the other hand, the shear modulus increases continuously from zero at the transition~\cite{Obukhov1995}.
Our result indicates that the system shows no 
fractal spatial dimension in the percolating cluster.
Therefore a mean-field approach, such as the coherent potential approximation
used in Ref.~\cite{Mao2010}, may be sufficient for the square lattice near its
rigidity threshold. Our results also verify the existence of a diverging
length scale $l^{*}\sim p^{-1}$, ignoring the slowly varying $\ln$ factor.  For
any given NNN bonds occupation probability $p$, lattices of sizes bigger than
$l^{*}$ are very likely to be rigid and lattices smaller than $l^{*}$ are very
likely to be floppy. This observation in square lattices is consistent with
the cutting argument on the isostatic length scale by Wyart~\cite{wyartthesis}.

Counting independent constraints is key in various systems that show a rigidity
transition, from jammed sphere packings~\cite{O'Hern2003} to 
rigidity percolation~\cite{Feng1984,Jacobs1995,Jacobs1996}. Mapping the
\emph{rigidity problem into the connectivity problem of random graphs} 
allows to draw inspiration from the vast body
of work on (random) graphs to gain insight in the rigidity problem. This
approach has led us to an exact expression for the probability that a square
lattice is made rigid by randomly adding next nearest neighbor bonds. We
speculate that this idea can be applied to a wider range of models for random
media, and advance our understanding of disordered materials.

\section{Acknowledgments}
It is a pleasure to thank Bryan Chen, Tom Lubensky, James Sethna, Gareth
Alexander, and Michael Schmiedeberg for useful discussions and suggestions.
This research is supported by the U.S. National Science Foundation through DMR-0804900
(XM), the U.S. Department of Energy, Office of Basic Energy Sciences, Division
of Materials Sciences and Engineering under Award DE-FG02-05ER46199 (WGE), and
by the Netherlands Organisation for Scientific Research (NWO) through a Veni
grant (WGE).
 
\section{Appendix: The infinite system size limit}
\label{sec:limit}
To obtain the thermodynamic limit $m,n\to\infty$, we follow Gilbert's strategy
of deriving a lower and upper bound to $\bigprob(n,n,\prob)$ and showing that these
are in close agreement~\cite{gilbert59}. The upper bound on $\bigprob(n,n,\prob)$ is given
by the lower bound on $1-\bigprob(n,n,\prob)$ that is set by the probability that at least one
of the $2n$ nodes is not connected to any other node. Denoting by $E_i$ the event that node
$i$ is not connected to any other node, which has probability $q^n$, we use a
Bonferroni inequality~\cite{fellerbook} to obtain
\begin{equation}
\label{bonferroni}
1-\bigprob(n,n,\prob)\geq P\left(\bigcup_{i} E_i\right) \geq \sum_i P(E_i)-\sum_{i<j}P(E_i\cap E_j)~.
\end{equation}
The first term on the right hand side equals $2nq^n$. The second term is 
of order $(nq^n)^2$ and can therefore be ignored as $n\to\infty$, so that the
upper bound becomes
\begin{equation}
\label{upperbound}
\bigprob(n,n,\prob)\leq 1-2nq^n\qquad\mathrm{as~}n\to\infty~.
\end{equation}

The lower bound is obtained directly from the recurrence relation~(\ref{recur2c}). We have,
using $\bigprob(k,l,\prob)\leq 1$ and writing $x=q^{m-2k}$,
\begin{eqnarray}
1-\bigprob(m,n,\prob)&=&{\sum_{k=1,l=0}^{m,n}}'\binom{m-1}{k-1}\binom{n}{l}\bigprob(k,l,\prob)q^{nk}x^l\nonumber\\
&\leq&\sum_{k=1,l=0}^{m,n}\binom{m-1}{k-1}\binom{n}{l}q^{nk}x^l-1\nonumber\\
&=&\sum_{k=1}^m\binom{m-1}{k-1}q^{nk}(1+x)^n-1\nonumber\\
&=&\sum_{k=1}^m\binom{m-1}{k-1}(q^{k}+q^{m-k})^n-1~,\nonumber\\
\label{binomexp}
\end{eqnarray}
where in the first line the prime indicates the term $(k,l)=(m,n)$ is to be excluded from the sum.
In the second line the sum is extended to include the term $(k,l)=(m,n)$
which is then corrected for by subtracting 1, and in the third line we used the
binomial expansion.  Now we can read off the leading terms: The terms for
$k=1,m-1,m$ together give $\bigprob(n,n,\prob)\geq 1-2nq^n-\order(n^2q^{2n-2})$,
while all other terms only give contributions of order $(nq^{n})^2$ or higher.
Hence we find convergence of this lower bound with the upper bound in
eq.~(\ref{upperbound}), and conclude that
\begin{equation}
\label{limitapp}
\bigprob(n,n,\prob)\to 1-2nq^n\qquad\mathrm{as~}n\to\infty~.
\end{equation}

This asymptotic form holds for finite $\prob$ and tells us how $\bigprob$
aproaches unity as bigger systems are considered. Note that the scaling of
$\pR$, defined as the value of $\prob$ where $\bigprob(n,n,\prob)=1/2$, does
not immediately follow form this because $\pR$ vanishes as $n\to\infty$. However,
we can obtain bounds on $\pR$ by considering where the bounds on $\bigprob$ cross $1/2$, as
long as we evaluate them keeping all orders of $nq^n$.
Equating the right hand side of eq.~(\ref{bonferroni}) to $1/2$ and solving for very large
$n$ we obtain $q^n=1/(2n)$, which gives a lower bound on $\pR$ of
\begin{equation}
\label{pRlowb}
\pR\geq\frac{\ln 2n}{n}\qquad\mathrm{as~}n\to\infty~.
\end{equation}
The upper bound on $\pR$ follows from equating the right hand side of eq.~(\ref{binomexp}) to $1/2$.
Numerically, we find again a solution of the form $q^n=1/(\beta n)$, but this time with $\beta\approx 4.93$, so
that
\begin{equation}
\label{pRbounds}
\frac{\ln 2n}{n}\leq\pR\lesssim\frac{\ln 4.93n}{n}~.
\end{equation}
From the result of the more elaborate proof by Pal\'asti~\cite{palasti63} one can derive that
$$
\pR=\frac{\ln(2n/\ln2)}{n}\approx\frac{\ln 2.89n}{n}~,
$$
which falls nicely within the bounds we derived by simpler means. In any case the leading order behavior is given by
\begin{equation}
\label{pRboundslimit}
\pR=\frac{\ln n}{n}+\order(1/n)\qquad\mathrm{as~}n\to\infty~.
\end{equation}

\end{document}